\documentclass[prb,aps,floatfix,twocolumn,amsmath,amssymb]{revtex4}
\usepackage{graphicx}
\usepackage{dcolumn}
\usepackage{bm}
\usepackage{longtable}
\usepackage[english,all,light]{draftcopy}

\newcommand{\pd}{\partial}
\newcommand{\mb}{\mathbf}
\usepackage{draftcopy}
\draftcopySetGrey{0.8}

\begin{document}

\title{High $Q$ Cavity Induced Fluxon Bunching in Inductively Coupled Josephson Junctions}

\author{S. Madsen}
\affiliation{Department of Chemistry, Aarhus University, DK-8000 Aarhus C, Denmark}

\author{N. Gr\o nbech-Jensen}
\affiliation{Department of Applied Science, University of California, Davis, California 95616}

\author{N. F. Pedersen}
\affiliation{Oersted$\cdot$DTU, Section of Electric Power Engineering, Technical University of Denmark, 2800 Kgs. Lyngby, Denmark}

\author{P. L. Christiansen}
\affiliation{Informatics and Mathematical Modeling and Department of Physics, Technical University of Denmark, 2800 Kgs. Lyngby, Denmark}

\date{\today}

\begin{abstract} 
We consider fluxon dynamics in a stack of inductively coupled long
Josephson junctions connected capacitively to a common resonant cavity
at one of the boundaries. We study, through theoretical and numerical analysis,
the possibility for the cavity to induce a transition from the
energetically favored state of spatially separated shuttling fluxons
in the different junctions to a high velocity, high energy state of identical
fluxon modes.
\end{abstract}

\maketitle

\section{Introduction} \label{intro}            
\setcounter{equation}{0}                        
THz emission from intrinsic Josephson junctions of the BSCCO type has received much attention recently. Several experiments have been reported \cite{exp1,exp2,exp3,exp4}, in which THz radiation emitted from BSCCO single crystals were observed. However in most cases the detected power is rather small, or the frequency is rather low, or the emitted radiation is detected indirectly on an on-chip detector. It has also been demonstrated that BSCCO can be considered a Josephson junction with ac Josephson effect even at frequencies as high as 2 THz \cite{exp5}. Recently a very convincing experiment was reported \cite{exp6} and it has attracted much focus and renewed experimental efforts.

Parallel to the experimental work there has been theoretical/numerical work on fluxon dynamics in in layered superconductors of the BSCCO type \cite{SBP,Solitons}. The calculations demonstrate that the best way to obtain THz radiation is by having in-phase motion of the fluxons in the different layers. This poses an interesting problem since the in-phase state of traveling fluxons is an energetically unfavorable state, and two fluxons of equal polarity will consequently repel each other. However, due to the disparity of wave speeds for in-phase and out-of-phase solutions, it has been shown \cite{NGJ_93,NGJ_94} that the energetically unfavorable in-phase state of traveling fluxons are be stable above the asymptotic speed of the out-of-phase mode. It has been assumed that the best way to obtain that is by having flux flow generated by a magnetic field applied parallel to the a-b plane, as studied theoretically in Ref. \onlinecite{NGJ_96}. However the successful experiment in Ref. \onlinecite{exp6} was done without a magnetic field, and it was suggested that an internal cavity (based on the so-called Fiske steps) played a major role in the THz generation. In this paper we study the fluxon modes in a stack of Josephson junctions interacting with an external cavity. We derive the conditions under which a large amount of current is induced in the cavity, and under which the external cavity may induce bunching of the Josephson junction fluxons.

\section{The Model} \label{TheModel}            

\begin{figure}
\centering
\includegraphics[width=3.25in]{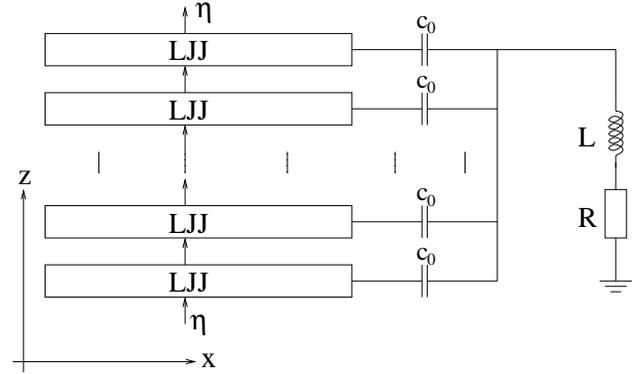}
\caption{The identical long Josephson junctions coupled to a cavity at $x=l$.} \label{Model}
\end{figure}

Assuming that all the junctions in the stack are identical, the equations for a stack of long Josephson junctions with $N+1$ superconducting layers and $N$ insulating layers can be written as \cite{SBP}
\begin{eqnarray}
\mathbf{J} = \mathbf{S}^{-1} \bm{\phi}_{xx}\ , \label{stackeq}
\end{eqnarray}

\noindent where the $i$'th element of $\bm{\phi}$, $\phi^i$, is the gauge invariant phase difference across insulating layer $i$. The $N\times N$ coupling matrix, $\mathbf{S}$, is given by (only non-zero elements are shown)
\begin{eqnarray}
\mathbf{S}=
\left( \begin{array}{ccccc}
1 & S &   &   & \\
S & 1 & S &   & \\
  & S & 1 & S & \\
  &   & \ddots & \ddots & \ddots
\end{array} \right)\ ,
\end{eqnarray}

\noindent with $S$ being the coupling parameter between the layers\cite{SBP}. The vector $\mathbf{J}$ has the components
\begin{eqnarray}
J^i = \phi^i_{tt} + \alpha \phi^i_t + \sin{\phi^i} - \eta\ , \label{current}
\end{eqnarray}

\noindent where dampng parameter $\alpha = 1/\sqrt{\beta}$ represents dissipation, and $\eta$ is the bias-current in the z-direction. Each component of $\mb{J}$ is a current in the $z$-direction.

Equations (\ref{stackeq})-(\ref{current}) have been written in normalized units. Space $x$ is normalized to the Josephson penetration depth, $\lambda_J=\sqrt{\hbar/2e\mu_0J^cd'}$, and time $t$ is normalized to the inverse plasma frequency $\omega_0^{-1}=\sqrt{\hbar c_J/2eJ^c}$, where $\mu_0$ is the vacuum permeability, $J^c$ is the critical current of the individual Josephson junctions, $d'$ is the effective thickness of the insulating layer, and $c_J$ is the capacitance of the individual junctions, see Refs. \onlinecite{SBP} and \onlinecite{Solitons} for details.

The model of the Josephson stack coupled to a series cavity is shown in Fig. \ref{Model}, where $L$ is the cavity inductance, $R$ is the cavity resistance, and $Nc_0$ is the total cavity capacitance. The boundary conditions for the phases can be written as\cite{MFP}
\begin{eqnarray}
\phi^i_x(0,t) &=& 0 \label{bc20}\ \ \ \ \ \textrm{and} \\
\phi^i_x(l,t) &=& \frac{\dot{q}}{N}-\frac{c}{N^2}\sum_{k=1}^N\left(\phi_{tt}^k(l,t)-\phi_{tt}^i(l,t)\right) \label{bc2L}\ ,
\end{eqnarray}

\noindent where $c=Nc_0/c_J$ is the normalized capacitance and $q$ is the normalized charge in the cavity. Defining $\Omega=1/(\omega_0\sqrt{NLc_0})$ to be the normalized cavity frequency and and $Q=\sqrt{L/(NR^2c_0)}$ as the quality factor, the linear cavity equation becomes
\begin{eqnarray}
\frac{d^2q}{dt^2} + \frac{\Omega}{Q} \frac{dq}{dt} + \Omega^2 q = \Omega^2 \frac{c}{N}\sum_{i=1}^N \phi^i_t(l,t)\ . \label{cavity}
\end{eqnarray}

\noindent For more details on these equations see Ref. \onlinecite{MFP}.

Two terms are present in Eq. (\ref{bc2L}). The first term couples the junctions to the cavity, by equally dividing the cavity current between the $N$ junctions. The second term represents a direct coupling between the junction through the capacitors $c_0$. In a real world situation, the junctions would be embedded in a resonator and couple through electro-magnetic radiation from the edges. The second term then models the radiation leaving one junction and ending up in another junction without being reflected by the cavity. This is clearly not very efficient due to the geometry of the stack. With an efficient cavity, the second term is therefore expected to be much smaller than the first term and may safely be neglected. This may be justified by numerical calculations. We thus choose to consider
\begin{eqnarray}
\phi^i_x(0,t) = 0 \ \ \ \ \textrm{and} \ \ \ \  \phi^i_x(l,t) = \frac{\dot{q}}{N} \label{bc}
\end{eqnarray}

\noindent as the boundary conditions for Eqs. (\ref{stackeq})-(\ref{current}) and (\ref{cavity}).

\section{Theoretical Analysis} \label{analytic} 
We analyze the system in Eqs. (\ref{stackeq})-(\ref{current}) and (\ref{cavity})-(\ref{bc}) in the case of weak inductive coupling where
\begin{eqnarray}
\mathbf{S}^{-1}=
\left( \begin{array}{ccccc}
1 & -S &   &   & \\
-S & 1 & -S &   & \\
  & -S & 1 & -S & \\
  &   & \ddots & \ddots & \ddots
\end{array} \right)\ , \label{Sinv}
\end{eqnarray}

\noindent valid to first order in $|S|$.

\subsection{Cavity Current}

The solution to the linear cavity equation with initial conditions $q(0)=0$ and $\dot{q}(0)=0$ is
\begin{eqnarray}
q(t)&=&\frac{e^{m_-t}}{m_--m_+}\int_0^t e^{-m_-t'}\frac{\Omega^2c}{N}\sum_{i=1}^N\phi^i_{t'}(l,t') dt' \nonumber \\
&-&\frac{e^{m_+t}}{m_--m_+}\int_0^t e^{-m_+t'}\frac{\Omega^2c}{N}\sum_{i=1}^N\phi^i_{t'}(l,t') dt' \ , \label{qParticular}
\end{eqnarray}

\noindent with $m_\pm\equiv-\Omega\left(1\pm i \sqrt{4Q^2-1}\right)/(2Q)$. The junction voltage at $x=l$, $\phi^i_t(l,t)$, thus generates the cavity charge. We look at the case where there is one fluxon in each junction and $\phi^i_t(l,t)$ then becomes a voltage pulse. To simplify the integrations, these pulses are approximated by delta functions, i.e.,
\begin{eqnarray}
\phi^i_t(l,t) = \sum_{n=0}^\infty A\delta\left(t-\tau^i-2\pi n/\omega^i\right) \label{deltaApprox}
\end{eqnarray}

\noindent approximating voltage pulses at $t=\tau^i+2\pi n/\omega^i$, $n=0,1,...$, $\omega^i$ is the fluxon shuttling frequency in junction $i$ and $\tau^i$ is the phase shift of junction $i$. Note, that since the present analysis is performed in the case of small $|S|$, all the $\delta$-functions will have approximate the same amplitude, $A$.

With this ansatz, the cavity current becomes
\begin{eqnarray}
\dot{q}(t)=\frac{\Omega^2c}{N}\sum_{i=1}^N\sum_{n=0}^\infty A H(\tilde{t}^i_n)\left[\cos\left(\frac{\Omega\sqrt{4Q^2-1}}{2Q}\tilde{t}^i_n\right)\right.\nonumber \\
-\left.\frac{1}{\sqrt{4Q^2-1}}\sin\left(\frac{\Omega\sqrt{4Q^2-1}}{2Q}\tilde{t}^i_n\right)\right]e^{-\frac{\Omega}{2Q}\tilde{t}_n^i}\ ,\label{genCur}
\end{eqnarray}

\noindent with $\tilde{t}^i_n\equiv t-\tau^i-2\pi n/\omega^i$ and where $H(t)$ is the Heaviside step function. Limiting the analysis to the case of a high $Q$ resonator, the steady state cavity current becomes
\begin{equation}
\dot{q}(t)=\frac{\Omega^2Ac}{N}\sum_{i=1}^N 
\frac{e^{\frac{\pi\Omega}{Q\omega^i}}\cos\left(\omega^i(t-\tau^i)+\varphi^i\right)}{\sqrt{1+e^{\frac{2\pi\Omega}{Q\omega^i}}-2e^{\frac{\pi\Omega}{Q\omega^i}}\cos\left(2\pi\Omega/\omega^i\right)}}\ \label{curGen}
\end{equation}
 
\noindent for $t\to\infty$. The phases, $\varphi^i$, are determined by
\refstepcounter{equation} \label{eqlabel}
\begin{equation*}
\cos\varphi^i = \frac{e^{\frac{\pi\Omega}{Q\omega^i}}-\cos\left(2\pi\Omega/\omega^i\right)}{\sqrt{1+e^{\frac{2\pi\Omega}{Q\omega^i}}-2e^{\frac{\pi\Omega}{Q\omega^i}}\cos\left(2\pi\Omega/\omega^i\right)}} \tag{\ref{eqlabel}a}\label{cphi}
\end{equation*}
\noindent and
\begin{equation*}
\sin\varphi^i = \frac{\sin\left(2\pi\Omega/\omega^i\right)}{\sqrt{1+e^{\frac{2\pi\Omega}{Q\omega^i}}-2e^{\frac{\pi\Omega}{Q\omega^i}}\cos\left(2\pi\Omega/\omega^i\right)}}\ . \tag{\ref{eqlabel}b}\label{sphi}
\end{equation*}

Only fluxons shuttling with the same frequency, $\omega^i\equiv\omega$ for $i=1,...,N$ will be considered. In this case, Eq. (\ref{curGen}) can be reduced to
\begin{eqnarray}
\dot{q}(t)=\epsilon\sum_{i=1}^N\cos\left(\omega\left(t-\tau^i\right)+\varphi\right)\ , \label{iphaselock}
\end{eqnarray}

\noindent with
\begin{eqnarray}
\epsilon \equiv \frac{\Omega^2Ac}{N\sqrt{1+e^{-\frac{2\pi\Omega}{Q\omega}}-2e^{-\frac{\pi\Omega}{Q\omega}}\cos\left(2\pi\Omega/\omega\right)}}\ , \label{epsilon}
\end{eqnarray}

\noindent and $\varphi^i\equiv\varphi$ for all $i$. Thus, the cavity current is very simple when the cavity has reached a steady state. Note that the amplitude of the cavity current for an in-phase mode ($\tau^i=\tau^j,\ i,j=1,...,N$) is $N\epsilon$. For an anti-phase mode ($\tau^i=\tau^{i+1}-(-1)^{i+1}\pi/\omega,\ i=1,...,N-1$) the amplitude is $\approx 0$ for $N$ even and $\approx \epsilon$ for $N$ odd.

Using Eq. (\ref{Sinv}), the Hamiltonian of the stack of weakly coupled Josephson junctions is
\begin{eqnarray}
H&=&\int_0^l\sum_{i=1}^N\Bigg(\frac{1}{2}(\phi^i_t)^2 + 1 - \cos\phi^i+\frac{1}{2}\phi^i_x\times \\
&&\left(\phi_x^i-S(1-\delta_{i,N})\phi_x^{i+1}-S(1-\delta_{i,1})\phi_x^{i-1}\right)\Bigg)dx \nonumber \ ,
\end{eqnarray}

\noindent with $\delta_{i,j}$ being the Kronecker delta function. Using Eqs. (\ref{stackeq})-(\ref{current}), (\ref{cavity}) and (\ref{Sinv}) the rate of change in energy is
\begin{eqnarray}
\frac{dH}{dt}&=&\sum_{i=1}^N\int_0^l\left(-\alpha\left(\phi_t^i\right)^2 + \eta\phi_t^i\right)dx+\Bigg[\sum_{i=1}^N\phi_x^i\times\label{dHdt} \\
&&\left(\phi_t^i-S(1-\delta_{i,N})\phi_t^{i+1}-S(1-\delta_{i,1})\phi_t^{i-1}\right)\Bigg]_0^l\nonumber \ .
\end{eqnarray} 

To determine the amplitude of the $\delta$-functions, we require that in the phase-locked state the energy-exchange of a ``collision'' with the boundary is the same for both a fluxon solution and the $\delta$-function approximation. This energy-exchange is given by the time-integral of the last term in Eq. (\ref{dHdt}),
\begin{eqnarray}
\Delta H_b = \sum_{i=1}^N\int_{t_1}^{t_2}\Bigg[\phi_x^i\Big(\phi_t^i&-&S(1-\delta_{i,N})\phi_t^{i+1}\label{DHb}\\
&-&S(1-\delta_{i,N})\phi_t^{i-1}\Big)\Bigg]_0^l dt\ ,\nonumber 
\end{eqnarray}

\noindent where $t_1$ and $t_2$ are taken such that they cover one collision with the boundary.

To model a fluxon collision with the boundary, the following profiles are used\cite{NGJ_thesis,Scott72}
\begin{equation}
\phi^i(x,t)=4\sigma^i\tan^{-1}\left(\frac{c_-}{u}\frac{\sinh\left(\left(t-\tau^i\right)u\gamma(u/c_-)/c_-\right)}{\cosh\left((x-l)\gamma(u/c_-)/c_-\right)}  \right)\ , \label{profile}
\end{equation}

\noindent with $\sigma^i=\pm 1$ determining the fluxon polarity, $\gamma(u)=1/\sqrt{1-u^2}$ being the Lorentz factor, and the lowest characteristic velocity $c_-^2\approx 1+2S\cos(\pi/(N+1))$ to first order in $|S|$. We take the same fluxon polarity in all junctions, thus $\sigma^i\equiv\sigma$ for all $i$.

Following Refs. \onlinecite{Delta,SGJ}, using Eqs. (\ref{bc}), (\ref{iphaselock}), and (\ref{profile}) in Eq. (\ref{DHb}) yields
\begin{eqnarray}
\Delta H_b^{f} &=& \frac{\sigma\zeta\epsilon}{N}\sum_{i=1}^N\Big[\left(1-S(2-\delta_{i,1}-\delta_{i,N})\right)\times \nonumber\\
&&\hspace*{1.4cm}\sum_{j=1}^N\cos\left(\omega(\tau^i-\tau^j)+\varphi\right)\Big] \label{deltaHb}
\end{eqnarray}

\noindent with
\begin{equation}
\zeta\equiv 4\pi\frac{\cosh\left(\frac{\omega c_-}{2u\gamma(u/c_-)}\cos^{-1}\left(2u^2/c_-^2-1\right)\right)}{\cosh\left(\frac{\pi\omega c_-}{2u\gamma(u/c_-)}\right)}\ , \label{zeta}
\end{equation}

\noindent and where the integration was carried out from $-\infty$ to $\infty$ for mathematical convenience.

Calculation of $\Delta H_b$ for the $\delta$-function approximation in Eq. (\ref{deltaApprox}) gives
\begin{eqnarray}
\Delta H_b^\delta &=& \frac{A\epsilon}{N}\sum_{i=1}^N\Big[\left(1-S(2-\delta_{i,1}-\delta_{i,N})\right)\times \nonumber\\
&&\hspace*{1.3cm}\sum_{j=1}^N\cos\left(\omega(\tau^i-\tau^j)+\varphi\right)\Big]\ .
\end{eqnarray}

Requiring $\Delta H_b^{f}=\Delta H_b^\delta$ determines the amplitude of the $\delta$-functions to
\begin{equation}
A = \sigma\zeta\ ,
\end{equation}

\noindent $\zeta$ being given by Eq. (\ref{zeta}) and $\sigma=\pm 1$.

\subsection{Current-Voltage Characteristics}
\noindent The asymptotic velocity, $u$, present in Eq. (\ref{profile}) may be determined similarly to what is outlined in Ref. \onlinecite{Delta}
\begin{eqnarray}
\frac{c_-}{u}\sinh\left(\frac{\pi u \gamma(u/c_-)}{2\omega c_-}\right) = \cosh\left(\frac{l\gamma(u/c_-)}{2c_-}\right)\ , \label{uAsymptotic}
\end{eqnarray}

\noindent when the fluxons are shuttling with frequency $\omega$. The conditions for a steady state require that the energy averaged over one period is zero, thus
\begin{eqnarray}
\Delta H = \int_{t_0-\frac{\pi}{\omega}}^{t_0+\frac{\pi}{\omega}} \frac{dH}{dt} dt = 0 \ .
\end{eqnarray}

\noindent Using Eq. (\ref{dHdt}), $\Delta H=0$ gives the condition
\begin{eqnarray}
\sigma\eta = \frac{\alpha}{IN}\int \int
\sum_{i=1}^N \left(\phi^i_t(x,t)\right)^2 dx dt - \frac{\Delta H_b}{2IN} \label{phaselock} \ ,
\end{eqnarray}

\noindent where $I$ is determined from\cite{Delta}
\begin{eqnarray}
\sinh\left(\frac{I\gamma(u/c_-)}{4\pi c_-}\right)=\frac{c_-}{u}\sinh\left(\frac{\pi u\gamma(u/c_-)}{2\omega c_-}\right)\ .
\end{eqnarray}

\noindent The current-voltage characteristics  in Eq. (\ref{phaselock}) include the phase $\varphi$, such that at a given bias current the system can adjust this phase together with the collision times, $\tau^i$, to satisfy condition (\ref{phaselock}) (if possible). The phase, $\varphi$, is related to the fluxon shuttling frequency, $\omega$, through Eqs. (\ref{cphi}) and (\ref{sphi}) and one may thus change the fluxon shuttling frequency by changing the bias current. From Eq. (\ref{phaselock}) $\eta$ is thus obtained as a function of $\omega$. In the numerical simulations in section IV we shall, inversely, obtain $\omega$ as function of $\eta$.

\subsection{Bunching}
To calculate the conditions for the cavity to induce bunching (in-phase motion), we consider a triangular fluxon configuration with one fluxon in each junction, modeled by $\tau^i=(-1)^ir/(2u)$. The interaction energy between the fluxons is first calculated by considering an infinite line with a lattice spacing of $r$, thus
\begin{equation}
\phi^i = 4\sigma\tan^{-1}e^{\frac{\gamma(u/c_-)}{c_-}\left(x-u\left(t-\frac{(-1)^ir}{2u}\right)\right)}\ .
\end{equation}

\noindent The interaction energy of this configuration is
\begin{eqnarray}
H_I = -S\sum_{i=1}^N\int_{-\infty}^\infty dx &\phi^i_x&\Big((1-\delta_{i,1})\phi^{i-1}_x\nonumber \\
&&+(1-\delta_{i,N})\phi^{i+1}_x\Big)\ ,
\end{eqnarray}

\noindent resulting in the well-known fluxon-fluxon force
\begin{eqnarray}
F_I &=& \int_{t_0-\frac{\pi}{\omega}}^{t_0+\frac{\pi}{\omega}} dt \frac{\pd H_I}{\pd r}\label{Fi} \\
&=& \frac{2\pi}{\omega}\frac{8 S(N-1)}{(c_-^2-u^2)\sinh\left(r/\sqrt{c_-^2-u^2}\right)}\times \nonumber \\
&& \left(1-\frac{r\cosh\left(r/\sqrt{c_-^2-u^2}\right)}{\sqrt{c_-^2-u^2}\sinh\left(r/\sqrt{c_-^2-u^2}\right)}\right) \ . \nonumber
\end{eqnarray}

\noindent The force on the fluxons from the boundary can be calculated from $\pd H/\pd r$ using $\phi^i_r=-(-1)^i\phi^i_t/2u$, valid for $0<x<l$, resulting in
\begin{eqnarray}
F_b &=& \int_{t_0-\frac{\pi}{\omega}}^{t_0+\frac{\pi}{\omega}} \frac{\pd H_b}{\pd r} dt \label{Fb}\\
 &=& -\frac{A\epsilon}{2uN}\sum_{i=1}^N\Big[(-1)^i\left(1-S(2-\delta_{i,1}-\delta_{i,N})\right)\times \nonumber\\
&&\hspace*{1.8cm}\sum_{j=1}^N\cos\left(\frac{\omega r}{2u}\left((-1)^i-(-1)^j\right)+\varphi\right)\Big]\ . \nonumber
\end{eqnarray}

The separation between the fluxons, $r$, may now be determined from the condition
\begin{eqnarray}
F_I + F_b = 0\ . \label{balance}
\end{eqnarray}

Solving Eq. (\ref{balance}) for $r$ enables one to calculate the current voltage characteristics, Eq. (\ref{phaselock}), and the cavity current, Eq. (\ref{iphaselock}), in the steady state for a triangular fluxon lattice.

\begin{figure}
\centering
\includegraphics[width=2in,angle=-90]{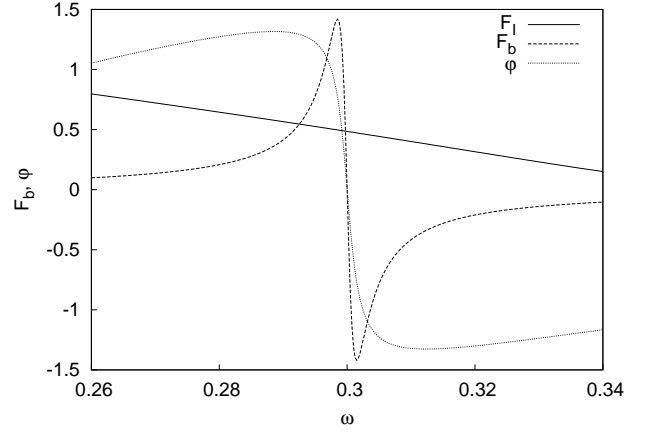}\vspace*{0.5cm}\\
\caption{$\varphi$ from Eqs. (\ref{cphi}) and (\ref{sphi}), $F_b$ from Eq. (\ref{Fb2}), and $F_I$ from Eq. (\ref{Fi}). $N=2,\ l=8, \alpha=0.1,\ Q=100,\ c=0.02,\ \Omega=0.3,\ S=-0.05,$ and $r=4$} \label{FigFb}
\end{figure}

For the case of only two junctions, Eq. (\ref{Fb}) reads
\begin{eqnarray}
F_b&=&(1-S)\frac{A\epsilon}{2u}\sin\varphi\sin\left(\frac{\omega r}{u}\right)\ . \label{Fb2}
\end{eqnarray}
In Fig. \ref{FigFb} we plot $\varphi$ from Eqs. (\ref{cphi}) and (\ref{sphi}) as well as $F_I$ from Eq. (\ref{Fi}) and $F_b$ from Eq. (\ref{Fb2}) as a function of $\omega$ at constant $r$. It is seen that when the system is above the resonance frequency, the force from the boundary is negative while the $F_I$ is positive, thus they may balance each other. Below the resonance frequency the two forces has the same sign and the only steady state solution must be the one where the fluxons move in anti-phase.

The perturbation to the current-voltage characteristic by the cavity is contained in the $\Delta H_b$-term in Eq. (\ref{phaselock}), given by Eq. (\ref{deltaHb}). In the case of anti-phase motion for two coupled junctions, this term will be zero. For three junctions, however, it will be non-zero. Eq. (\ref{Fi}) and (\ref{Fb}) will have the same direction for $\omega<\Omega$ resulting in anti-phase motion below the resonance frequency. This suggests that only for an odd number of junctions, we may observe an area of negative differential resistance in the current-voltage characteristics.

\section{Numerical Simulations} 
\begin{figure}
\centering
\includegraphics[width=2in,angle=-90]{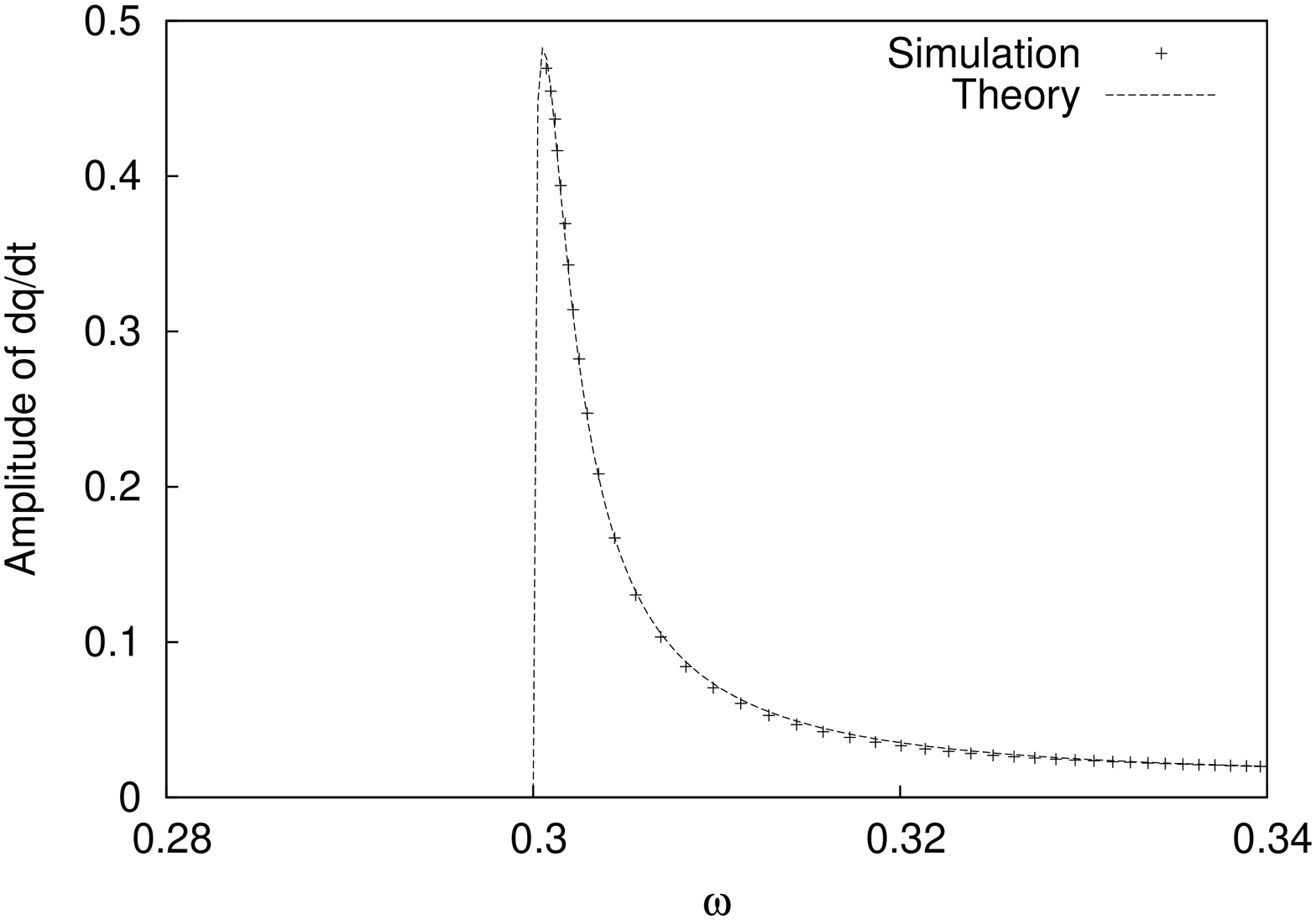}\vspace*{1cm}\\
\includegraphics[width=2in,angle=-90]{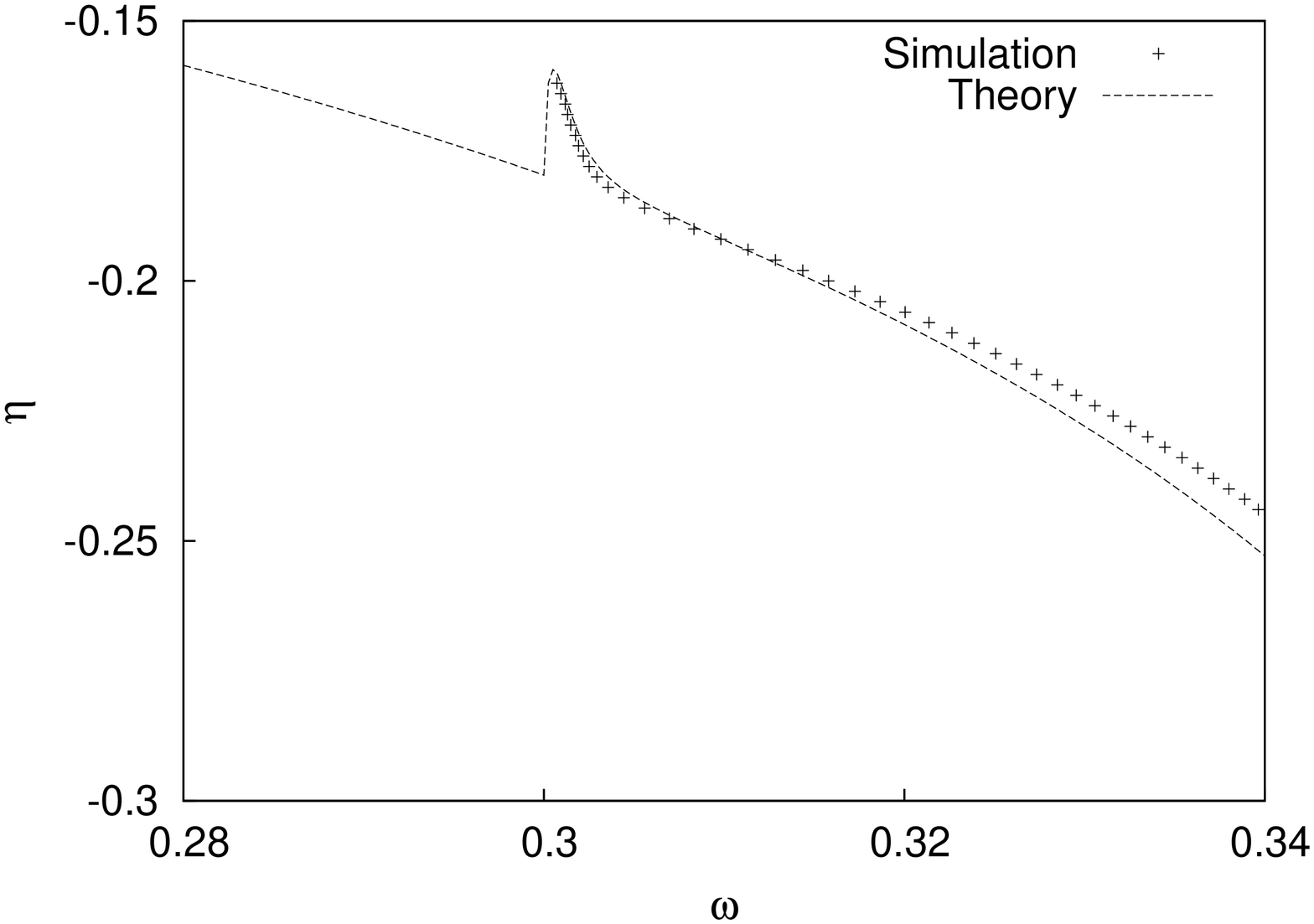}\vspace*{1cm}\\\
\includegraphics[width=2in,angle=-90]{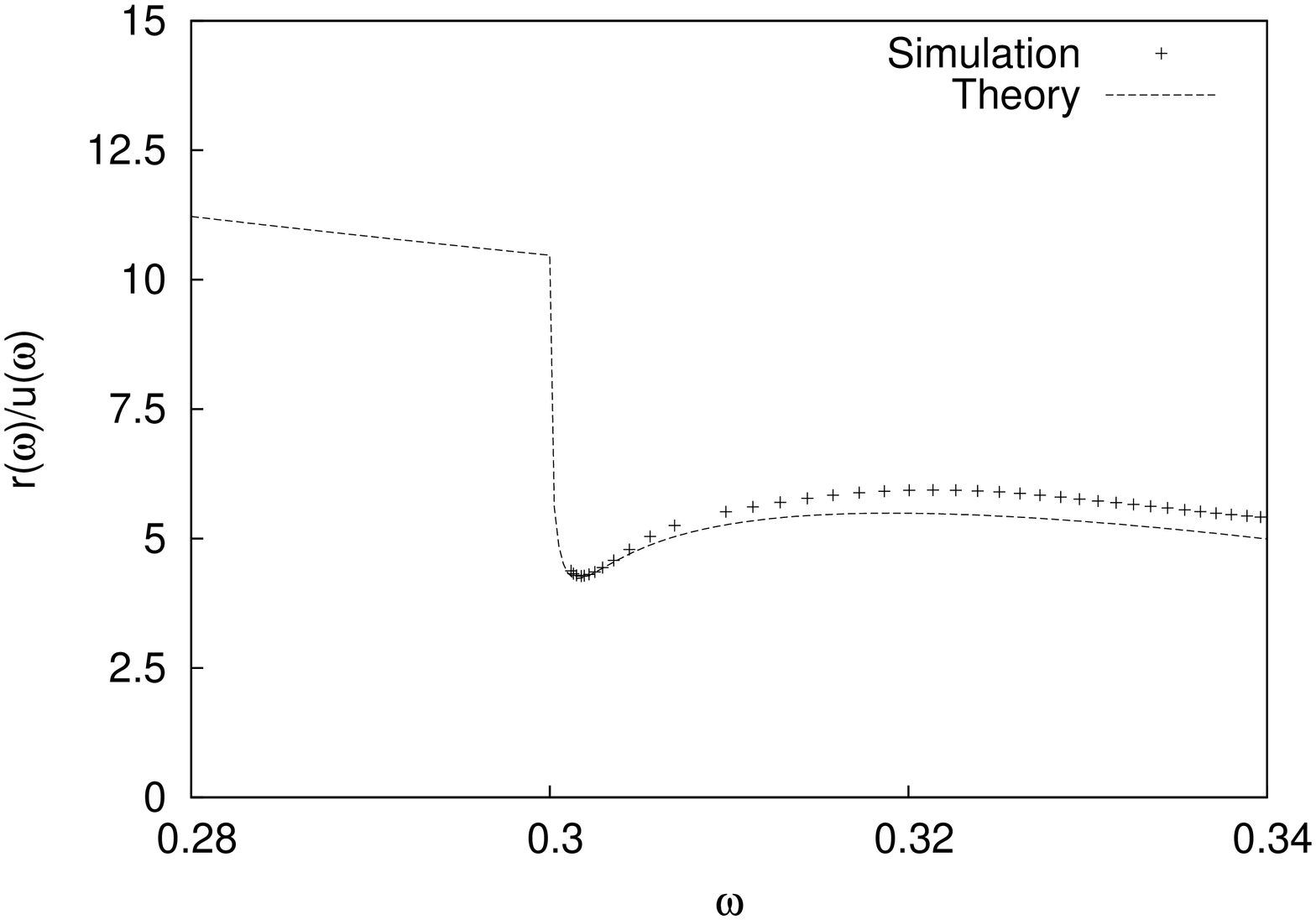}\vspace*{0.5cm}\\
\caption{Amplitude of $\dot{q}$ (top), current-voltage (middle), and $r/u$ (bottom). $N=2,\ l=8, \alpha=0.1,\ Q=100,\ c=0.02,\ \Omega=0.3,$ and $S=-0.05$. $\omega$ is the fluxon shuttling frequency, controlled by the bias current ($\eta$) in experiments.} \label{L8}
\end{figure}

\begin{figure}
\centering
\includegraphics[width=2in,angle=-90]{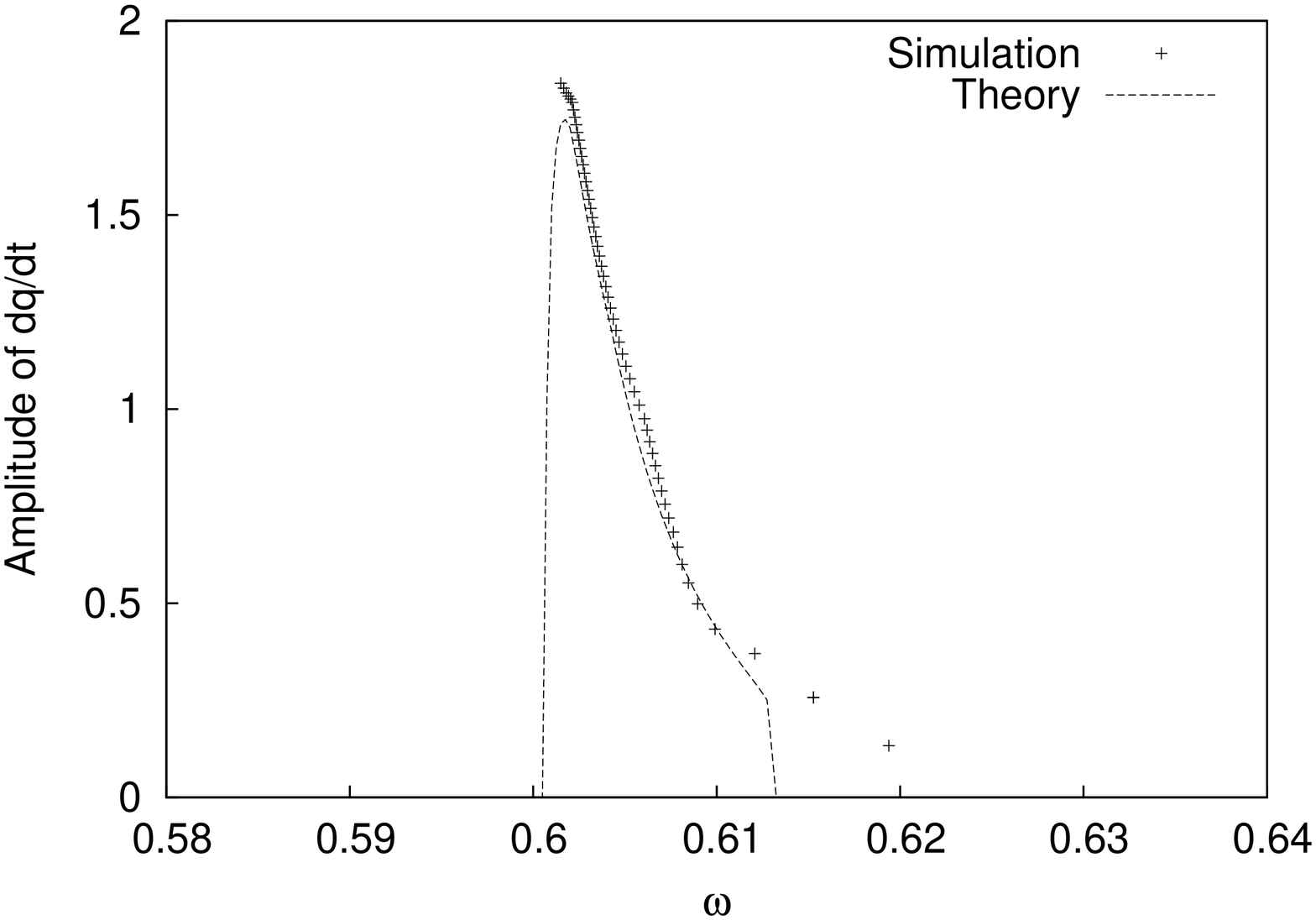}\vspace*{1cm}\\
\includegraphics[width=2in,angle=-90]{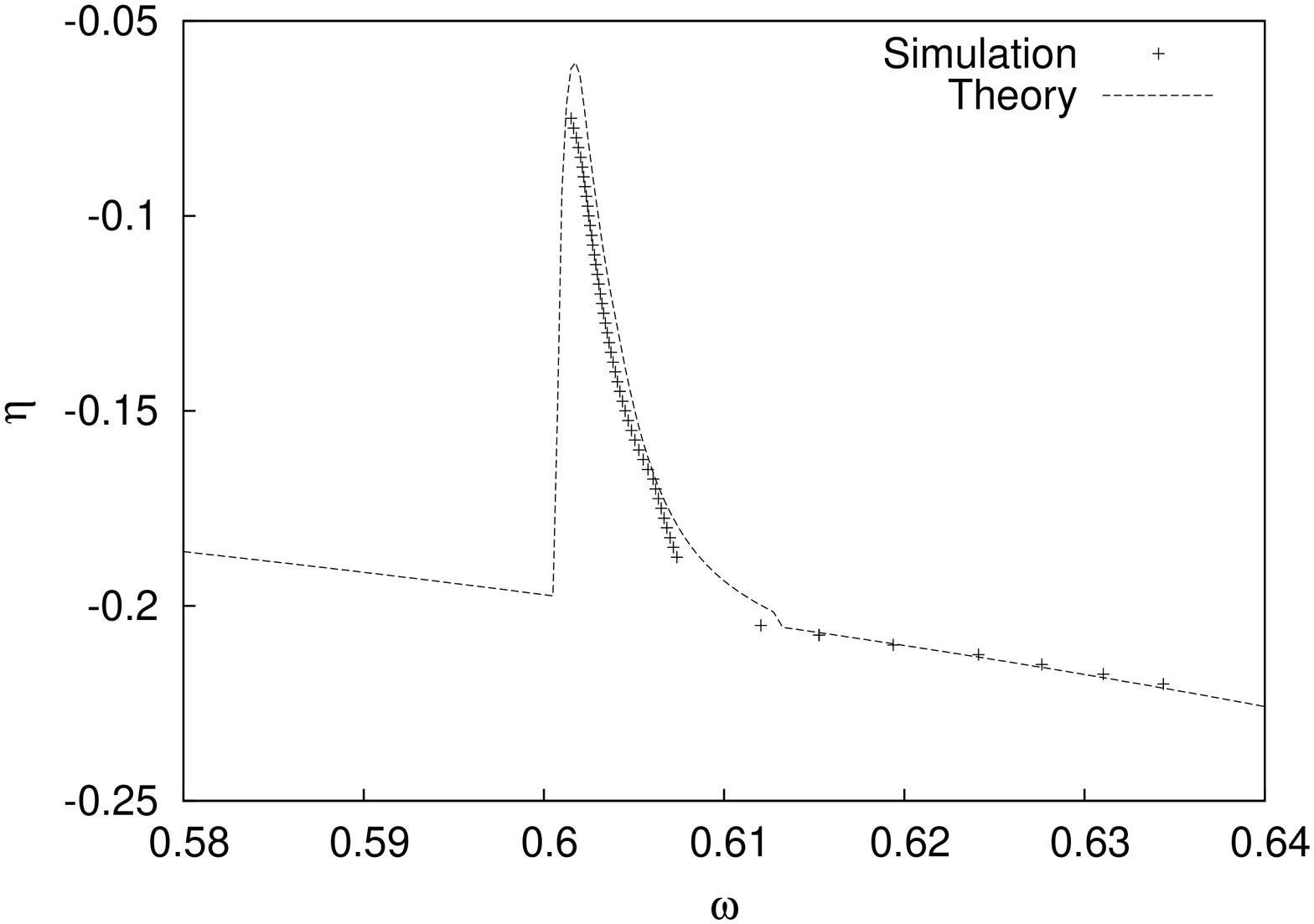}\vspace*{1cm}\\\
\includegraphics[width=2in,angle=-90]{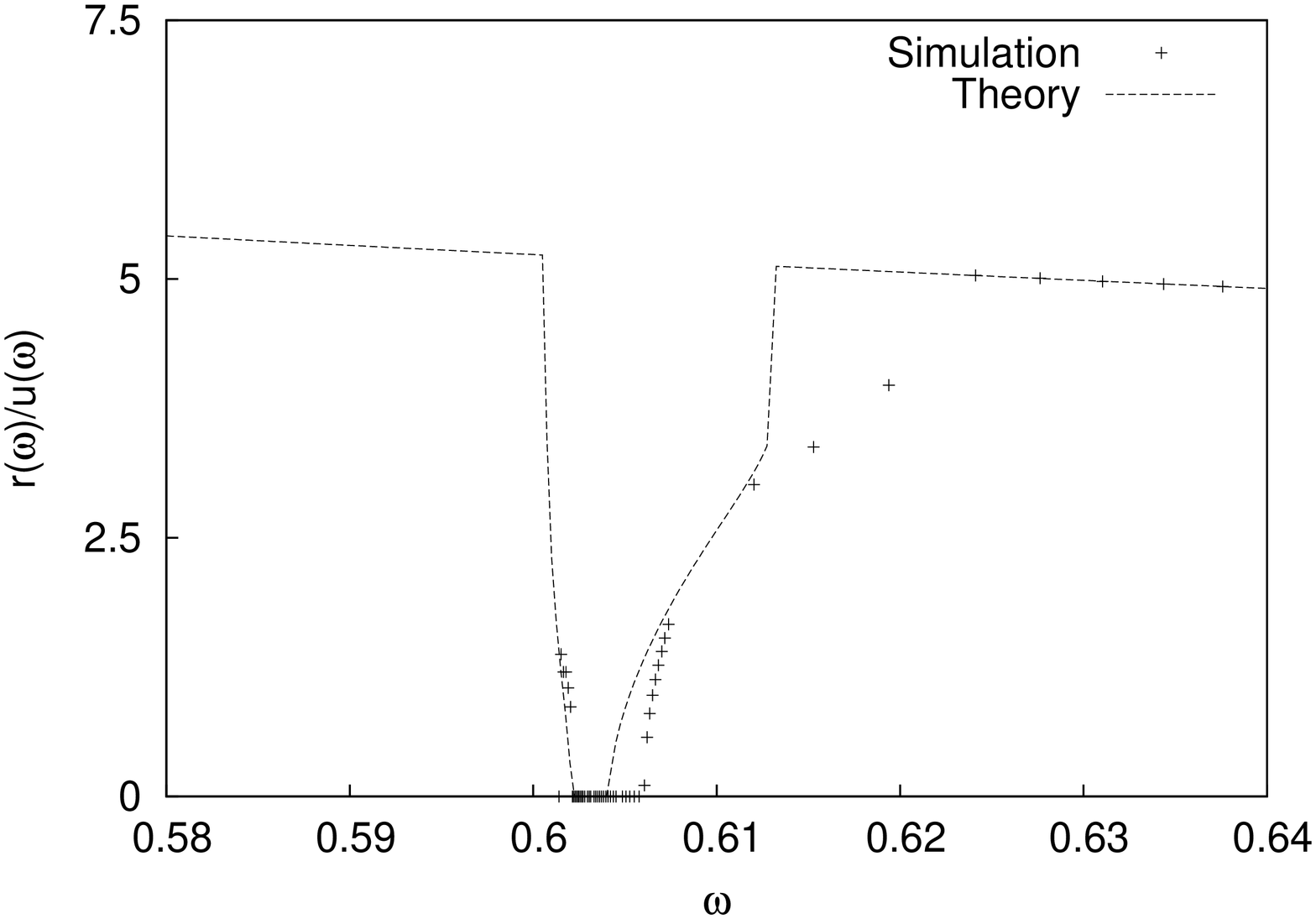}\vspace*{0.5cm}\\
\caption{Amplitude of $\dot{q}$ (top), current-voltage (middle), and $r/u$ (bottom). $N=2,\ l=4, \alpha=0.1,\ Q=100,\ c=0.02,\ \Omega=0.6,$ and $S=-0.05$. $\omega$ is the fluxon shuttling frequency, controlled by the bias current ($\eta$) in experiments.} \label{L4}
\end{figure}

\begin{figure}
\centering
\includegraphics[width=2.5in,angle=-90]{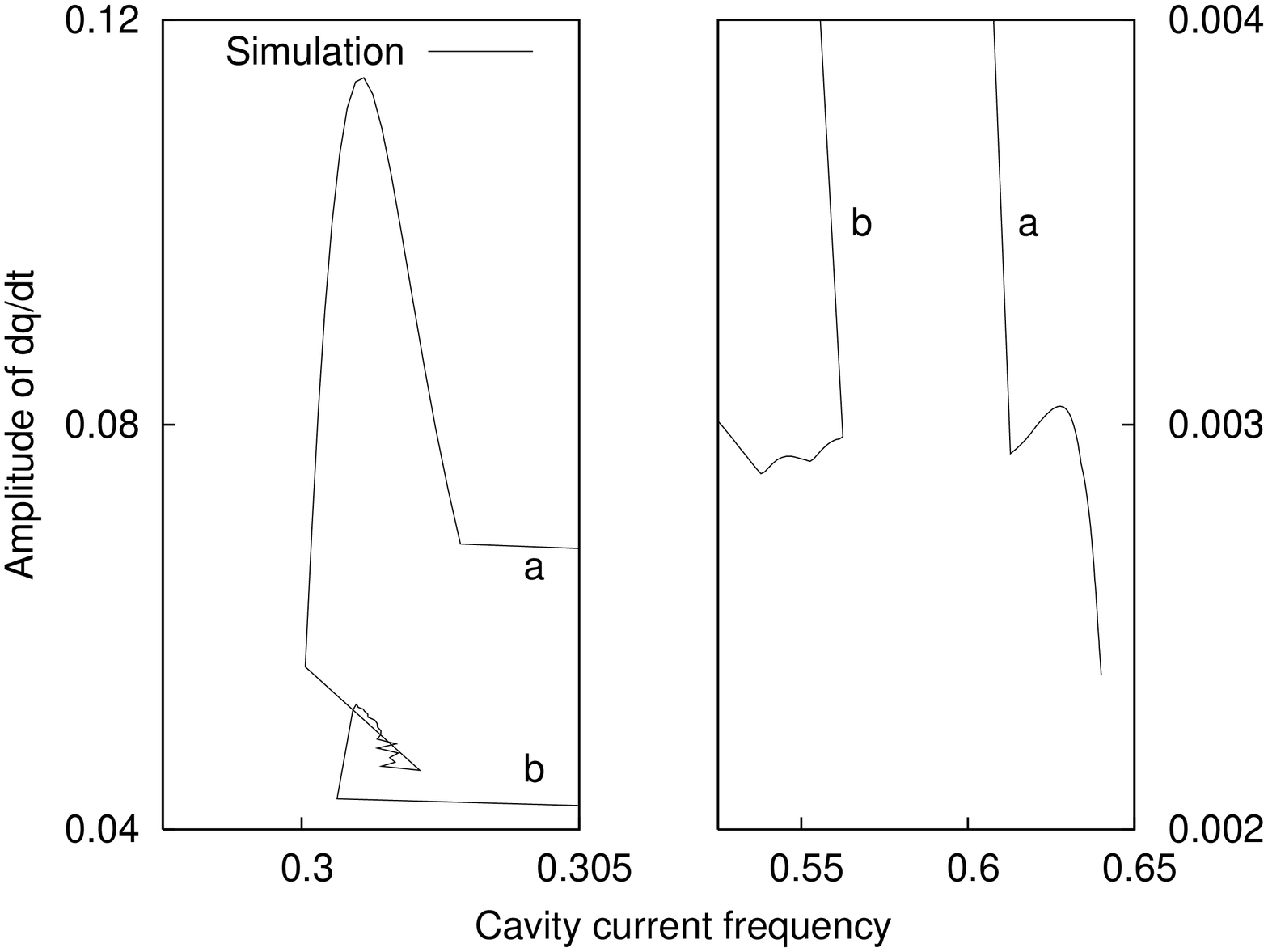}\vspace*{1cm}\\
\includegraphics[width=2.5in,angle=-90]{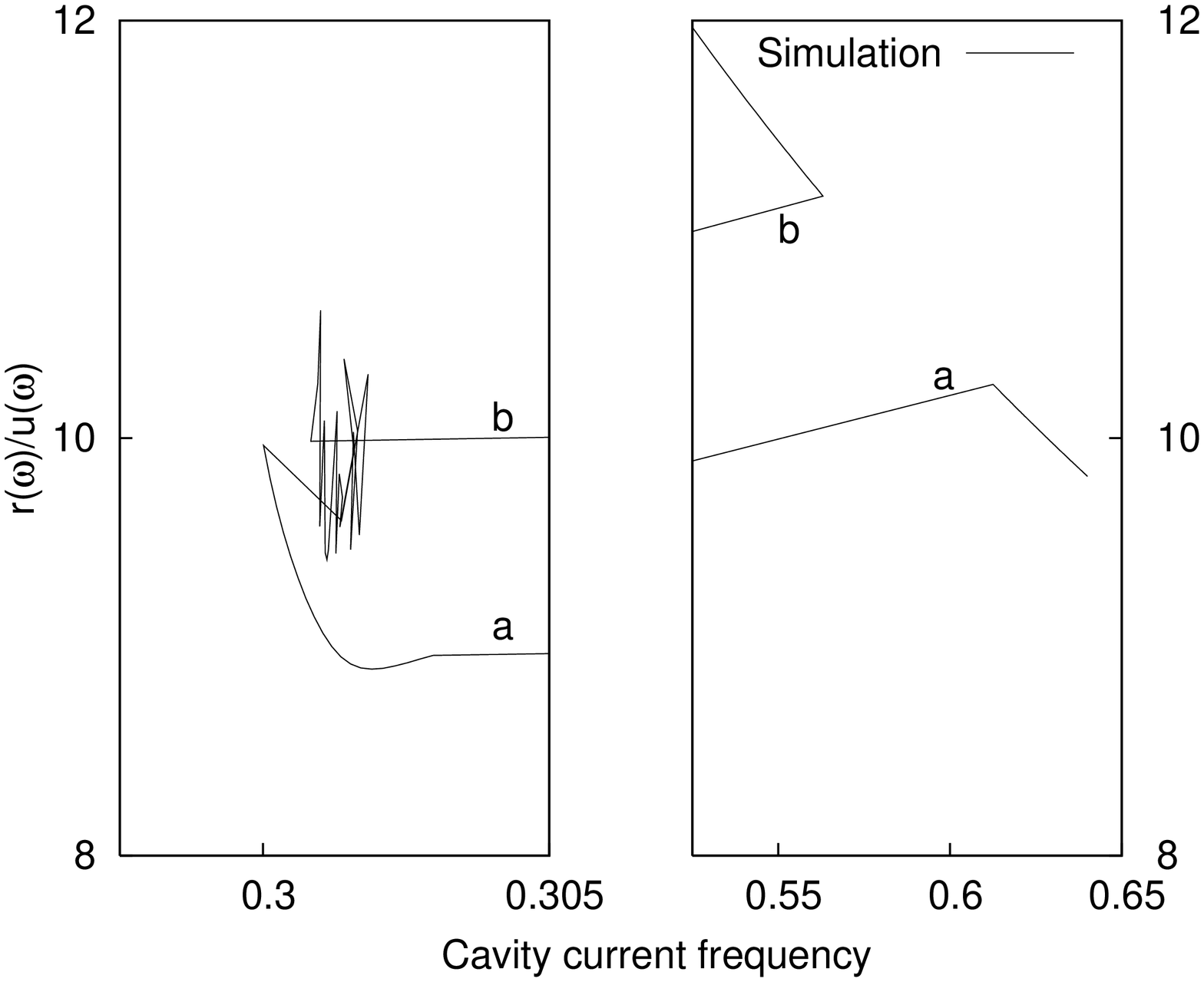}\\
\caption{Amplitude of $\dot{q}$ (top) and $r/u$ (bottom) as a function of the cavity current frequency. $N=2,\ l=8, \alpha=0.1,\ Q=100,\ c=0.02,\ \Omega=0.3,$ and $S=-0.45$.  $\omega$ is the fluxon shuttling frequency, controlled by the bias current ($\eta$) in experiments. The simulation was started at high bias current resulting in high cavity frequency and then the bias current was lowered resulting in a lowering of the cavity current frequency. At line a the system switches from a high cavity current frequency to a low cavity current frequency state. At point b the system switches from a low cavity current frequency to a high cavity current frequency state.} \label{L8HighS}
\end{figure}

Numerical simulations of the full non-linear Eqs. (\ref{stackeq})-(\ref{current}) with boundary conditions (\ref{cavity}) and (\ref{bc}) has been done using second order finite differences for the spatial derivatives and a 5th order Runge-Kutta method with adaptive step size for the temporal integration\cite{nr}. The spatial resolution was kept at $0.01$ for all considered systems. The initial fluxon configuration had one fluxon in each junction, each moving in anti-phase with the one in the neighboring junction(s). The system was integrated until a stabilized cavity current was obtained or 20000 time units had passed. The stable cavity current, $\dot{q}$, and the individual voltages at $x=l$, $\phi_t(l,t)$, was analyzed using interpolation and FFT\cite{nr} to determine the most significant frequency in the power spectrum which is used as the cavity current frequency and the fluxon shuttling frequency, $\omega$.

The difference in collision times, $\delta\tau=r/u$, can be calculated directly from the simulation using $\phi^i_t(l,t)$. It may also be calculated analytically using Eqs. (\ref{uAsymptotic}) and (\ref{balance}), where the latter Eq. was solved numerically for $r$. Sometimes there were multiple solutions, $r_i$, and we have chosen $r=min\{r_i\}$. When no solution was found in the interval, we used $r/u=\pi/\omega$ corresponding to anti-phase motion. The amplitude of the stabilized cavity current can be determined using a simple line search in the $\dot{q}$ data and compared to the amplitude of Eq. (\ref{iphaselock}) using the value of $r$ obtained from Eq. (\ref{balance}). The frequency versus applied bias current can be compared to Eq. (\ref{phaselock}), again using $r$ obtained from Eq. (\ref{balance}).

When a steady state is found, we observe that $\omega_{cavity}=n\omega_{fluxon}$ with $n=1$ or $2$, with $n=2$ giving a very low cavity current and therefore no significant difference from the unperturbed system. For high bias currents, the system was observed to be in the $n=2$ state and a switching to $n=1$ occurred when the system came close to the resonance in the current voltage characteristic. Below resonance frequency, the system again switched to the $n=2$ state. To compare the analytical and numerical results, we only show the numerical simulation points where a steady state was reached for the $n=1$ case.

In Figs. \ref{L8} and \ref{L4} we have used Eqs. (\ref{iphaselock}), (\ref{phaselock}), (\ref{Fi}), and (\ref{Fb})/(\ref{Fb2}) for the analytical results (shown with dashed lines).

Fig. \ref{L8} shows the results on the system with $N=2,\ l=8, \alpha=0.1,\ Q=100,\ c=0.02,\ \Omega=0.3,$ and $S=-0.05$. Well above the resonance frequency we get very little current in the cavity. As the shuttling frequency approaches the cavity frequency, the cavity current increases and reaches a maximum at near cavity frequency but suddently drops to near zero slightly above the cavity frequency. Note that all numerical results are only shown for frequency values larger than the resonance frequency. This in general agreement with the findings in Ref. \onlinecite{xx} where only oscillators with frequencies higher than the resonance frequency can be syncronised. In addition, the forces shown in Fig. \ref{FigFb} are seen to be directed in opposite directions for frequencies larger than the resonance frequency and in the same direction for frequencies smaller than the resonance frequency. The fluxon-fluxon distance will thus decrease only if the shuttling frequency is larger than the resonance frequency. The corresponding current-voltage characteristic thus shows a deviation from the case without a cavity only above the resonance frequency. The fluxon separation shows a similar behavior. Exactly at the resonance frequency, where the cavity current is at its maximum, the system exhibit anti-phase motion due to the boundary force being zero and we find the system to be in the $n=2$ state in the numerical simulations. Slightly above resonance frequency, the separation has a minimum and then it increases until it reaches at maximum at some point and then it decreases again. The cavity current is thus largest slightly above resonance frequency, since the boundary force is zero exactly at resonance frequency, resulting in anti-phase fluxon motion.

Fig. \ref{L4} shows the corresponding results on the system with $N=2,\ l=4, \alpha=0.1,\ Q=100,\ c=0.02,\ \Omega=0.6,$ and $S=-0.05$. The general behavior is the same as the one in Fig. \ref{L8}, except for the fluxon-fluxon separation. Below and at resonance frequency, we again see anti-phase behavior. Slightly above the resonance frequency, we see that the fluxon-fluxon separation has decreased to zero, i.e. the system has switched to a bunched state. At higher fluxon shuttling frequency, the fluxon are separated at some distance $r$ and at some point this distance become so great that we again see anti-phase motion.

The minor discrepancies observed in Figs. \ref{L8} and \ref{L4} between theory and numerical experiment are primarily caused by the two core assumptions in the perturbation analysis; namely the rigid collective coordinate approximation for the fluxon, and the idealized treatment of the fluxon reflection at the boundaries of the junction. Among the approximations inherent to these assumptions are omission of phonons and the change in fluxon dynamics during reflections. We notice, however, that the agreement between theory and simulations is very good, as can be seen in the figures.

The rather weak force induced by the cavity on the fluxons can only be used to obtain bunching in the weakly coupled case. It is, however, essential that the fluxons do not move in perfect anti-phase in order to induce current into the cavity. The top plot of Fig. \ref{L8HighS} the amplitude of the cavity current is shown for a simulation with similar parameters as Fig. \ref{L8} but with a much higher inductive coupling, $S=-0.45$, approaching the case of intrinsic junctions. The simulation was started with a high bias current, resulting in a high fluxon-shuttling frequency and the bias current was gradually lowered resulting in lower fluxon shuttling frequencies. Eq. (\ref{iphaselock}) gives zero cavity current in the case of a perfect anti-phase mode. In the numerical simulations, however, we do not find zero cavity current, but rather that the system is in the $n=2$ state, i.e. the cavity is oscillating with twice the fluxon shuttling frequency. As the fluxon shuttling frequency is lowered, the cavity current increases enough to slightly break the anti-phase motion and the cavity starts to oscillate at the fluxon shuttling frequency, seen in Fig. \ref{L8HighS} by following the line marked with 'a' from the high frequency part to the low frequency part. As the fluxon shuttling frequency is lowered still, the amplitude of the cavity current increases and the fluxon separation, $r/u$, decreases. Near the resonance the cavity current gets smaller and the fluxon separation start to increase again. At some point, the fluxons start to behave erratic, meaning that we can not find a definitive value of $r$ in the simulations and thus we do not obtain steady state motion. The part of the curve near $\omega=0.302$ where the value of $r/u$ oscillates heavily shows this. At some point the system again switches to the anti-phase motion with the cavity oscillating at twice the fluxon frequency, seen by following the 'b'-line from the low frequency part to the high frequency part of the figure. We have not been able to determine if the erratic behaviour near $\omega=0.302$ is due to a too short simulation time before we give up finding a steady state of if this is the 'true' behaviour of the system.

\section{Conclusion}      
We have analytically calculated the cavity current and the current-voltage relation for $N$ weakly inductively coupled stacked Josephson junctions coupled to a resonance cavity. We have shown that the cavity introduces a force between the natively repulsive fluxons which may be used to obtain bunching in the weakly coupled case. The effect is strongest for short junction, where the boundaries have larger influence. Our simple analysis show overall good agreement with numerical simulations. In the case of high inductive coupling our perturbation results deviate from the simulations but the overall picture is still consistent with the theory.

\section{Acknowledgments} 

SM and NFP would like to acknowledge the STVF framework program ``New Superconductors: Mechanisms, processes and products'' for financial support. SM would also like to acknowledge financial support from the Lundbeck Foundation.


\bibliographystyle{unsrt}

\end{document}